\documentclass[onecolumn,trackchanges]{aastex631}
\usepackage{amsmath}

\usepackage{xcolor}

\usepackage{ulem}
\shorttitle{circular polarization of a hybrid energy distribution of electrons}
\shortauthors{Cheng et al.}
\graphicspath{{./}{figures/}}
\begin{document}

\title{Synchrotron Circular Polarization in Gamma-Ray Burst Prompt Optical Emission: Relativistic Thermal Electron Contribution}

\correspondingauthor{Jirong Mao}
\email{jirongmao@mail.ynao.ac.cn}

\author[0009-0004-3324-8421]{Kangfa Cheng}
\affiliation{Guangxi Key Laboratory for Relativistic Astrophysics, School of Physical Science and Technology, Guangxi University \\
Nanning 530004, People’s Republic of China}
\affiliation{School of Physics and Electronic Information, Guangxi Minzu University \\
Nanning 530006, People’s Republic of China}

\author{Jirong Mao}
\affiliation{Yunnan Observatories, Chinese Academy of Sciences \\ 
Kunming, 650011, People’s Republic of China}
\affiliation{Key Laboratory for the Structure and Evolution of Celestial Objects, Chinese Academy of Science \\
Kunming, 650011, People’s Republic of China }
\affiliation{Center for Astronomical Mega-Science, Chinese Academy of Science \\
20A Datun Road, Chaoyang District, Beijing 100012, People’s Republic of China}

\author{Xiaohong Zhao}
\affiliation{Yunnan Observatories, Chinese Academy of Sciences \\ 
Kunming, 650011, People’s Republic of China}
\affiliation{Key Laboratory for the Structure and Evolution of Celestial Objects, Chinese Academy of Science \\
Kunming, 650011, People’s Republic of China }
\affiliation{Center for Astronomical Mega-Science, Chinese Academy of Science \\
20A Datun Road, Chaoyang District, Beijing 100012, People’s Republic of China}

\author{Hongbang Liu}
\affiliation{Guangxi Key Laboratory for Relativistic Astrophysics, School of Physical Science and Technology, Guangxi University \\
Nanning 530004, People’s Republic of China}

\author{Zhegeng Chen}
\affiliation{Guangxi Key Laboratory for Relativistic Astrophysics, School of Physical Science and Technology, Guangxi University \\
Nanning 530004, People’s Republic of China}
\affiliation{School of Physics and Electronic Information, Guangxi Minzu University \\
Nanning 530006, People’s Republic of China}

\author{Gaojin Yu}
\affiliation{School of Physics and Electronic Information, Guangxi Minzu University \\
Nanning 530006, People’s Republic of China}

\author{Zhifu Chen}
\affiliation{School of Physics and Electronic Information, Guangxi Minzu University \\
Nanning 530006, People’s Republic of China}



\begin{abstract}
Synchrotron circular polarization of a non-thermal power-law electron distribution in gamma-ray bursts (GRBs) has been studied. However, some numerical simulations have shown that the resulting distribution of electrons is a combination of a thermal component and a non-thermal power-law component. In this paper, we investigate synchrotron circular polarization using such a hybrid energy distribution of relativistic thermal and nonthermal electrons within a globally toroidal magnetic field in GRB prompt optical emission.  Our results show that compared to the solely nonthermal electron model, the synchrotron circular polarization degree (PD) in the hybrid electron model can vary widely in the optical band, depending on different parameters. The lower the electron temperature, the higher the circular PD. The time-averaged circular PD in the hybrid electron model can be higher than $\sim 1\%$ when the electron temperature is as low as $\sim 10^{10}$ K, while in the solely nonthermal electron model is usually lower than $\sim 1\%$. We further calculate the radiative transfer of the circular and linear polarization in the optical band. Our results show that both of the circular and linear PDs decrease with the increase of optical depth, but the linear PDs decline faster than the circular PDs.
To further examine the physical mechanisms of both radiation and particle acceleration, we expect that 
instruments will be capable of measuring the circular polarization of GRB prompt optical emission in the future.
\end{abstract}



\section{Introduction} \label{sec:intro}
Gamma-ray bursts (GRBs) are the most intense high-energy astrophysical explosive phenomena in the universe. Although significant progress has been made in the GRB research field, several fundamental issues remain unclear,
such as the radiation mechanism, the magnetic field (MF) structure in the radiation region, and the jet structure. The polarization of GRBs can serve as a probe for investigating these issues. 
Synchrotron radiation is the leading mechanism in GRB prompt emission, such that the study of synchrotron polarization is important. Synchrotron polarization consists of linear and circular parts.
The synchrotron linear polarization in GRB prompt emission has been widely studied (e.g. \citealt{Granot+konigl+2003, Nakar+etal+2003, Toma+etal+2009, Cheng+etal+2020, cheng+etal+2024b, Cheng+etal+2024, Lan+etal+2019, Lan+etal+2020, Lan+etal+2021(2), Lan+etal+2021, Li+etal+2024, Tuo+etal+2024, Gill+etal+2020, Gill+etal+2021, Gill+Granot+2021, Gill+Granot+2024}), while the synchrotron circular polarization in GRB prompt phase is not well noticed.
\cite{Sazonov+1969} has calculated the intrinsic synchrotron circular polarization degree (PD), and it is related to the electron Lorentz factor (LF) as $P_{cir} \propto \gamma^{-1}$ when we consider single electron case.
It can be found that the synchrotron circular PD is generally produced by relativistic electrons with relatively low energy. Currently, the research on the circular polarization of GRBs is primarily focused on the afterglow phase. \citep{Matsumiya+etal+2003, Sagiv+etal+2004, Nava+etal+2014, Nava+etal+2016, Mao+Wang+2017}. Some circular polarization observations in the afterglow of GRBs have been reported. \cite{Wiersema+etal+2012} found the optical afterglow of GRB 091018 with circular PD $P_{\rm{cir}} <0.15\%$. \cite{Wiersema+etal+2014} reported that the optical afterglow of GRB 121024A was measured with circular PD $P_{\rm{cir}}=0.61\pm 0.13\%$, and it was accompanied with evolution.

However, the research on synchrotron circular polarization in GRB prompt optical emission remains scarce. A population of GRBs has been observed with prompt optical emission, such as GRBs 990123, 041219A, 060111B, 080319B, 130427A, 160625B, and 180325A. The origin of the prompt optical emission of GRBs is still under debate. 
Several models, such as the reverse shock emission \citep{Sari+Piran+1999a}, the internal shock emission \citep{Meszaros+Rees+1999, Wei+2007}, and the synchrotron self-Compton emission from the high-latitude region \citep{Panaitescu+Kumar+2007}, have been proposed. GRB 990123 is the first GRB observed with prompt optical flash \citep{Sari+Piran+1999b}. GRB 080319B is the first GRB  confirmed to be bright enough to be seen with the naked eye \citep{Bloom+etal+2009, Cenko+etal+2010}. In particular, GRB 160625B was observed with prompt optical flash, and the optical flash was detected with linear polarization evolution \citep{Troja+etal+2017}. Therefore, in principle, in addition to the linear polarization,  circular polarization can be also detectable in the prompt optical emission. We expect that the future polarization instruments will be capable of detecting the circular polarization of prompt optical emission. 
In this paper, we will study the synchrotron circular polarization in the GRB prompt optical emission.

Synchrotron radiation is usually assumed to be produced by relativistic nonthermal electrons following a power-law distribution. The relativistic nonthermal electrons can originate from internal shocks (e.g. \citealt{Rees+Meszaros+1994, Paczynski+Xu+1994}) or the dissipation of MF energy (e.g. \citealt{Usov+1992, Thompson+1994, Spruit+etal+2001, Vlahakis+2003, Zhang+Yan+2011}). However, recent particle-in-cell (PIC) simulations of baryon-dominated relativistic shocks have found that only a small fraction ($\sim 10\%$) of energy dissipated into the power-law distribution of electrons, and the majority of the energy is deposited into a Maxwellian distribution of electrons \citep{Spitkovsky+2008a, Spitkovsky+2008b, Giannios+Spitkovsky+2009, Martins+etal+2009, Sironi+Spitkovsky+2009}. The electrons with Maxwellian distribution are thermal components. The PIC simulations of the Poynting-flux-dominated case 
have also found that the electron distribution evolves gradually from an initial thermal distribution to a combination of a thermal distribution and a nonthermal distribution \citep{Guo+etal+2014, Guo+etal+2015}. 
These physical results of particle acceleration research provide a hint that a realistic electron distribution is a hybrid energy distribution of relativistic thermal and nonthermal electrons. \cite{Mao+Wang+2018} calculated the intrinsic synchrotron linear polarization by considering such a hybrid electron distribution model. Further, \cite{cheng+etal+2024b} considered jet geometric effects and calculated the synchrotron linear polarization in GRB prompt emission under this hybrid electron model. \cite{Mao+etal+2018} (hereafter"Mao18") calculated the synchrotron circular polarization and the polarization radiative transfer by the hybrid electron model. However, the circular PD calculated in Mao18 is intrinsic polarization and has not considered any geometric effects of GRB jet. The jet geometric effects and MF structure could lead to a significant change in the observed PDs. Hence, in order to better explain the observations, we will take the geometric effects of the jet and MF structure into account to calculate the circular PD of the GRB prompt optical emission in the hybrid electron model in this paper.

Considering the absorption of polarized photons by the medium surrounding the synchrotron source, polarization radiative transfer should be considered in the optical and radio bands in the polarization measurements. \cite{Sazonov+1969} and \cite{Jones+ODell+1977} have studied the synchrotron polarization radiative transfer of relativistic nonthermal electrons. Mao18 calculated the synchrotron polarization radiative transfer of the hybrid electron distribution of thermal and nonthermal electrons. As mentioned in the preceding paragraph, the calculations in Mao18 did not take into account jet geometric effects and MF structure. Therefore, we will consider these two effects to calculate the synchrotron polarization radiative transfer in the GRB prompt optical emission. In such a case, both linear and circular polarization processes are considered simultaneously.  

This paper is organized as follows. Section \ref{model} describes the models in this article, including the electron distribution model, jet model, and MF model. Section \ref{sec:cal-pol} shows the calculations of synchrotron circular PD in the optical band in GRB prompt emission, including instantaneous and time-averaged circular PDs in a single pulse. Section \ref{sec:transfer} calculate the synchrotron circular and linear PDs by taking the radiation transfer into account in GRB prompt optical emission. The summary and discussions are shown in Section \ref{sec: conclusion}.

\section{Models} \label{model}

\subsection{Electron composition: a hybrid energy distribution of relativistic thermal and nonthermal electrons} \label{sec:dis-ele}
The PIC simulations of both baryon-dominated relativistic shocks and Poynting-flux-dominated outflow have indicated that the resulting electron distribution is a combination of relativistic thermal and non-thermal components.
In this paper, we adopt this electron distribution to calculate the synchrotron circular PD of GRB prompt optical emission. 
The electron distribution is divided into two segments: a thermal distribution plus a nonthermal distribution. Electrons with LFs below a conjunctive LF ($\gamma_{th}$) follow a Maxwell distribution, while electrons with LFs above the conjunctive LF follow a power-law distribution. The electron distribution is presented as \citep{Giannios+Spitkovsky+2009, Mao+Wang+2018, cheng+etal+2024b}

\begin{equation} \label{ele-dis}
N_{e} (\gamma) =
\begin{cases}
N_{0} \gamma^{2} \exp(-\gamma/\Theta)/2\Theta^{3}   & (\rm{for \quad \gamma \leq \gamma_{th}}) \\
N_{0} \gamma_{th}^{2} \exp(-\gamma_{th}/\Theta)(\gamma/\gamma_{th})^{-p}/2\Theta^{3}  & (\rm{for \quad \gamma >\gamma_{th}})
\end{cases}
\end{equation} 

where $p$ is the power-law index of the nonthermal electron energy distribution, $N_{0}$ is a dimensionless normalization constant, $\gamma$ is the electron LF, and $\Theta= kT_{e}/m_{e}c^{2}$ is the characteristic temperature. $k$ is Boltzmann constant, $T_{e}$ is the temperature of the relativistic thermal electrons, $m_e$ is the electron mass, and $c$ is the speed of light. 
To facilitate our study of the impact of electron distribution on the synchrotron circular PD, we define a parameter of nonthermal energy fraction $f$. This parameter represents the proportion of the energy of nonthermal electrons to the total energy of electrons. It is defined as \citep{Giannios+Spitkovsky+2009, Mao+Wang+2018, cheng+etal+2024b}

\begin{equation}
f=\frac{\int_{\gamma_{th}}^{\infty} \gamma N_{e}(\gamma,\Theta) d\gamma}{\int_{\gamma_{min}}^{\infty}\gamma N_{e}(\gamma,\Theta)d\gamma}.
\end{equation} \label{frac}

where $\gamma_{\rm{min}}$ is the minimum LF of the electrons, and in this paper we take $\gamma_{\rm{min}}=3$ below which the synchrotron approximation becomes invalid. The nonthermal energy fraction $f$ is dependent on the conjunctive LF $\gamma_{th}$ and the temperature $T_e$ of the thermal electrons. 

\subsection{Jet and Magnetic field model} 
The jet structure of GRBs remains uncertain. It can be classified into two structures: a uniform top-hat jet and a structured jet. The structured jets are primarily hypothesized to have two forms: one where the angular energy follows a power law distribution, and the other where it follows a Gaussian distribution. For simplicity, we take the uniform top-hat jet in our calculations.  

The MF configuration in the radiation area of GRB prompt emission is unclear. The small-scale random field and the large-scale ordered field are two candidate MF topologies in GRB prompt emission. These two MF models have different origins. The small-scale random field originates from the Weibel instability \citep{Gruzinov+Waxman+1999, Medvedev+Loeb+1999} or the kinetic turbulence \citep{2011ApJ...731...26M, 2013ApJ...776...17M, Mao+Wang+2017}, while the large-scale ordered MF is generated from the central objects of GRB \citep{Spruit+2001}. \cite{Matsumiya+etal+2003} pointed out that the tangled MF model can not generate synchrotron circular polarization. Here, we consider the large-scale ordered MF model in this paper. 
The large-scale ordered MF generally contains two components: radial and toroidal components. The strength of radial component decreases as $B' \propto R^{-2}$, while the toroidal component decreases as $B' \propto R^{-1}$ \citep{Spruit+2001}. The radiation area in GRB  prompt emission is usually at a large radius, thus the MF should be dominated by the toroidal component in the emission region. The MF strength in the jet comoving frame is presented as \citep{Spruit+2001, Uhm+Zhang+2014, cheng+etal+2024b}
\begin{equation}
B'(t')=B'_0 [\frac{R(t')}{R_0}]^{-1},
\end{equation}
where $B'_0$ is the initial MF strength in the jet comoving frame, $R_0$ is the initial radius, and $R(t')=R_0+\beta c\Gamma t'$ is the jet radius. $\Gamma$ and $\beta$ are the bulk LF and the dimensionless velocity of the jet, respectively. $t'= t_{\rm{obs}}\delta/(1+z)$ is the jet comoving time, 
where $t_{\rm{obs}}$ is the observed time in the laboratory frame, $\delta$ is the Doppler factor, and $z$ is the redshift.

\section{Calculations of the circular polarization} \label{sec:cal-pol}
In this paper, we adopt the hybrid electron distribution model to calculate the synchrotron circular PD of GRB prompt optical emission. Our aim is to present the circular polarization characteristics under this model, with the hope of validating this model in the future through polarization measurements.
The circular polarization of a single relativistic electron can be written as \citep{Sazonov+1969}
 \begin{equation} \label{pi-single}
 \pi_{c,s}=\frac{4 \cos \alpha' [x K_{1/3} (x) +\int_{x}^{\infty} K_{1/3} (\xi)d\xi] }{3 \gamma \sin \alpha' [x\int_{x}^{\infty}K_{5/3}(\xi)d\xi]}
 \end{equation}
where $K_{5/3}(\xi)$ and $K_{1/3}(x)$  are Bessel functions, $x=\nu'/\nu'_{c}$, $\nu'_c=\frac{3q_{e}B'\sin\alpha'}{4\pi m_e c}\gamma_{e}^{'2}$, and $\alpha'$ is the pitch angle. Note that the primed symbols denote physical quantities in the jet comoving frame. For an electron population with a distribution of $N_{e}(\gamma)$, the total circular PD is \citep{Sazonov+1969}
 \begin{equation} \label{pi-pop}
\pi_{c,p}=\frac{4 \int_{\gamma_{\rm{min}}}^{\gamma_{\rm{max}}} \cos \alpha' [N_{e}(\gamma)/\gamma]  [x K_{1/3} (x) +\int_{x}^{\infty} K_{1/3} (\xi)d\xi] d\gamma}{3 \int_{\gamma_{\rm{min}}}^{\gamma_{\rm{max}}} \sin \alpha' N_{e}(\gamma) [x\int_{x}^{\infty}K_{5/3}(\xi)d\xi] d\gamma}
 \end{equation}
Considering the jet geometric effects, the circular PD in the prompt optical emission can be calculated as
\begin{equation} \label{pi-0}
\begin{aligned}
\Pi_{C,0} (\nu)  &=4 \int_{0}^{(1+q)^{2}y_{j}} g(y) dy \int_{-\Delta \phi(y)}^{\Delta \phi(y)} d\phi  \int_{\gamma_{\rm{min}}}^{\gamma_{\rm{max}}}  [N_{e}(\gamma)/\gamma] [x K_{1/3} (x) +\int_{x}^{\infty} K_{1/3} (\xi)d\xi]\\
&\times B'(t') \cos \alpha' d\gamma  [3 \int_{0}^{(1+q)^{2}y_{j}} g(y) dy \int_{-\Delta \phi(y)}^{\Delta \phi(y)} d\phi \\
&\times  \int_{\gamma_{\rm{min}}}^{\gamma_{\rm{max}}}  N_{e}(\gamma) B'(t')  \sin \alpha' [ x\int_{x}^{\infty}K_{5/3}(\xi)d\xi] d\gamma]^{-1}.
\end{aligned}
\end{equation}

 Some variables are defined here: $y \equiv (\Gamma \theta)^{2}$,  $y_{j} \equiv (\Gamma
\theta_{j})^{2}$, and $q \equiv \theta_{v} / \theta_{j}$, where $\theta_{j}$ is the jet opening angle, and $\theta_{v}$ is the viewing angle. For the time-resolve PD, $g(y)=(1+y)^{-3}$, while for the time-averaged PD, $g(y)=(1+y)^{-2}$ \citep{Nakar+etal+2003}. By taking the electron distribution Equation \ref {ele-dis} into Equation \ref{pi-0}, we can calculate the instantaneous and time-averaged PD of GRB prompt emission. Other variables are as follows \citep{Granot+2003, Granot+konigl+2003, Granot+Taylor+2005, Toma+etal+2009}:
\begin{equation}
\sin \alpha' = \left[ \left(\frac{1-y}{1+y}\right)^{2} + \frac{4y}{(1+y)^{2}} \frac{(s-\cos \phi)^{2}}{(1+s^{2}-2s \cos \phi)} \right]^{1/2},
\end{equation}
%
%
\begin{equation}
\Delta \phi(y) =
\begin{cases}
0,    \qquad \qquad \qquad \qquad \rm{for} \; q>1  \; \rm{and}  \;  y<(1-q)^2 y_{j}     \\
\pi,  \qquad \qquad \qquad \qquad \rm{for} \; q<1  \; \rm{and}  \;  y<(1-q)^2 y_{j}   \\
\cos^{-1} \left[\frac{(q^{2}-1)y_{j}+y}{2q \sqrt{y_{j}y}}\right]   \qquad \qquad \qquad \qquad \quad \rm{otherwise}.
\end{cases}
\end{equation}
where $s=\theta / \theta_{v}$.

\subsection{Instantaneous circular polarization in GRB prompt optical emission} \label{subsec:time-resolve}
\begin{figure}[ht!]
\plotone{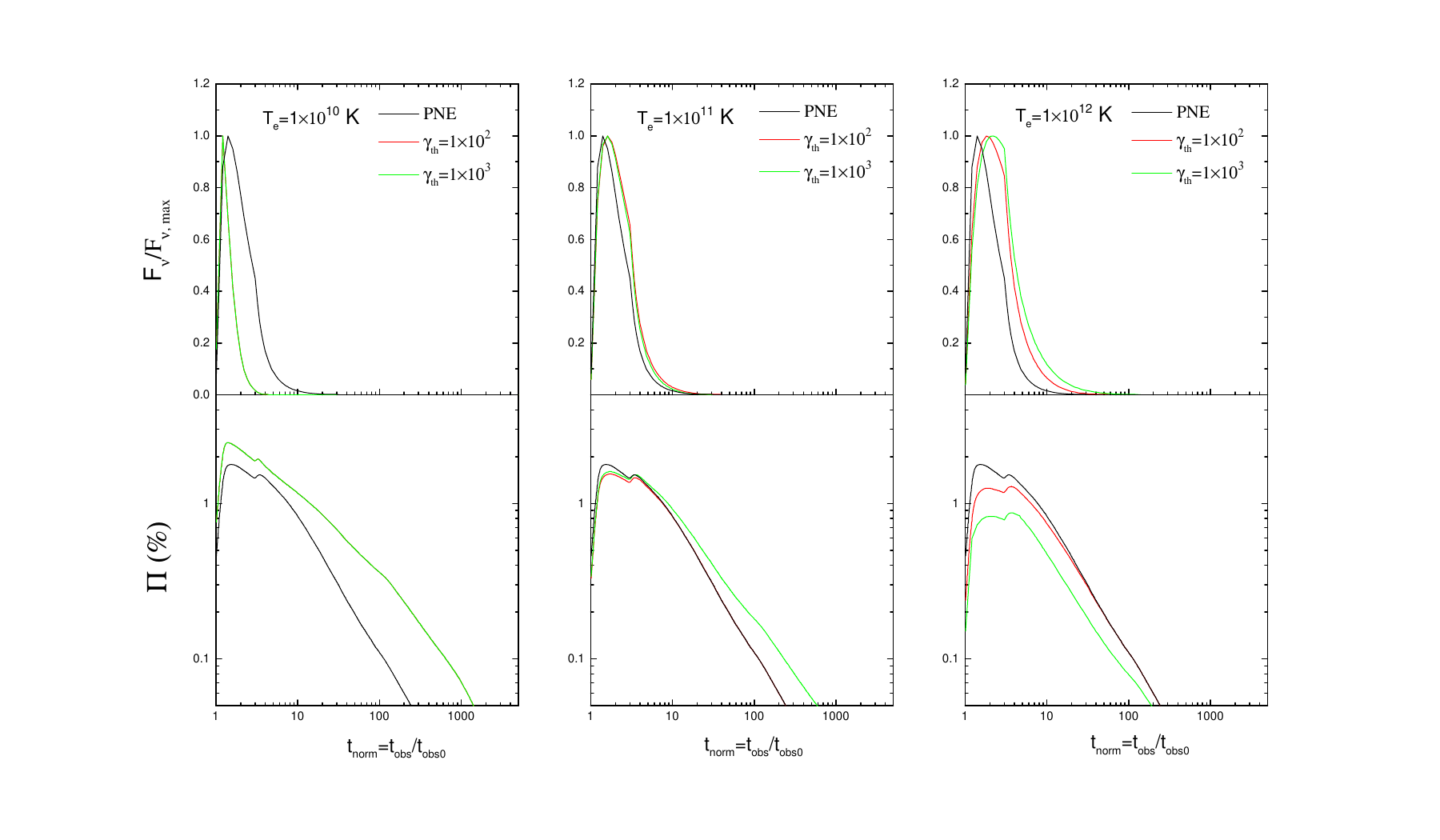}
\caption{Instantaneous circular PDs with different temperatures ($T_e$) and conjunctive LFs ($\gamma_{th}$) in the optical band ($6\times 10^{14}$Hz). We take $q=0.8$, $\Gamma=300$, and $B=100 G$ in these calculations. The upper panels display the normalized light curves (scaled to the maximum flux), while the bottom panels show the corresponding instantaneous circular PDs. PNE represents the synchrotron radiation produced by the pure nonthermal electrons. \label{fig:res-cgth}}
\end{figure}

We calculate the instantaneous circular PDs in a single pulse in the optical band ($6\times 10^{14}$Hz). For example, the optical band is within the detected wavebands of the optical polarimeter MOPTOP \citep{Shrestha+etal+2020}. Instantaneous circular PDs with different temperatures ($T_e$) and conjunctive LFs ($\gamma_{th}$) are shown in Figure \ref{fig:res-cgth}. The synchrotron polarization produced by the pure nonthermal electrons is also shown in the figures for comparison, and the model is marked as "PNE".  The time $t_{\rm{norm}}$ in the X-axis is normalized to the starting time $t_{\rm{obs0}}$ of the emission, where $t_{\rm{obs0}}=R_{0}(1+z)[1-\beta]/(\beta c)$ and $t_{\rm{obs0}}=R_{0}(1+z)[1-\beta \cos(\theta_{v}-\theta_{j})]/(\beta c)$ for the on-beaming and off-beaming cases, respectively. As is shown in Figure \ref{fig:res-cgth}, the circular PD evolution for all polarization curves is similar. The circular PD increases with time in the rising phase of the light curve and decreases in the decaying phase. Thus the evolution of circular polarization should be dominated by geometric effects, rather than difference radiation cases.
The circular PDs with the same conjunctive LFs for different temperatures display significant differences when compared to the left, middle, and right panels.  The lower temperatures the higher circular PDs. Moreover, the circular PDs with the same temperatures for different conjunctive LFs also display differences. 
The circular PDs obtained from the hybrid electron energy distribution could be higher than those obtained from the PNE model for the case of the electron temperature as low as $\sim 10^{10}$ K and $\gamma_{th} \gtrsim 10^{2}$, and the highest circular PD can reach $\sim 2\%$.

\subsection{Time-averaged circular polarization in GRB prompt optical emissin}\label{subsec:averaged}
In addition to the instantaneous polarization, the time-averaged polarization is also important. In order to impose stricter constraints on the GRB model, we also calculate the time average circular PDs. We calculate the time-averaged circular PDs in a single pulse with different parameters in the optical band in this section.

\subsubsection{Time-averaged circular PD compared with the local circular PD}
\begin{figure}[ht!]
\plotone{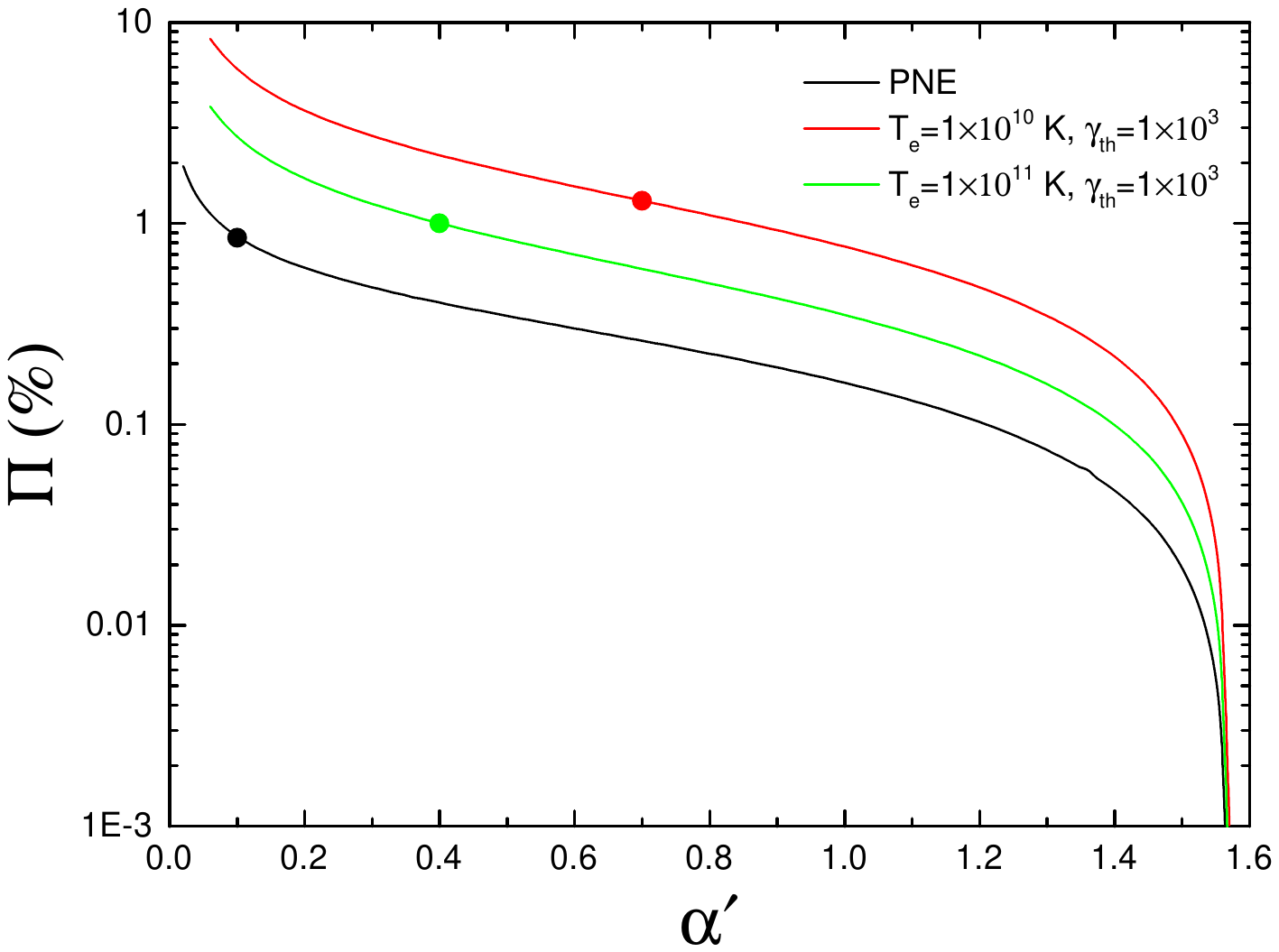}
\caption{The local circular PDs profile with different pitch angles in the optical band ($6\times 10^{14}$Hz). Data points in different colors represent the time-averaged circular PDs under different parameter sets. The "PNE" model represents the synchrotron radiation produced by the pure nonthermal electrons. We take $q=0.8$ for time-averaged polarization in this group of calculations. \label{fig:pi-alpha}}
\end{figure}

To investigate the effect of jet geometry on the circular polarization, we calculate both the local circular PDs (without jet geometric effects) and the time-averaged circular PDs (incorporating jet geometric effects) of GRB prompt emission and compare them in Figure \ref{fig:pi-alpha}. The local and time-averaged circular PDs can be calculated by using the Equations \ref{pi-pop} and \ref{pi-0}, respectively. As is shown in Figure \ref{fig:pi-alpha}, the larger the pitch angle $\alpha'$ the lower the local circular PDs. 
By comparing the time-averaged circular PDs to the local ones, we find that the time-averaged circular PDs are lower than the highest local PDs under different parameter sets. The reduction in time-averaged circular PDs compared to local maxima results from the integration across the jet's geometry structure, where the partial cancellation caused by geometric symmetry suppresses net circular polarization.

\subsubsection{Time-averaged circular PD profiles with viewing angles}

\begin{figure}[ht!]
\plotone{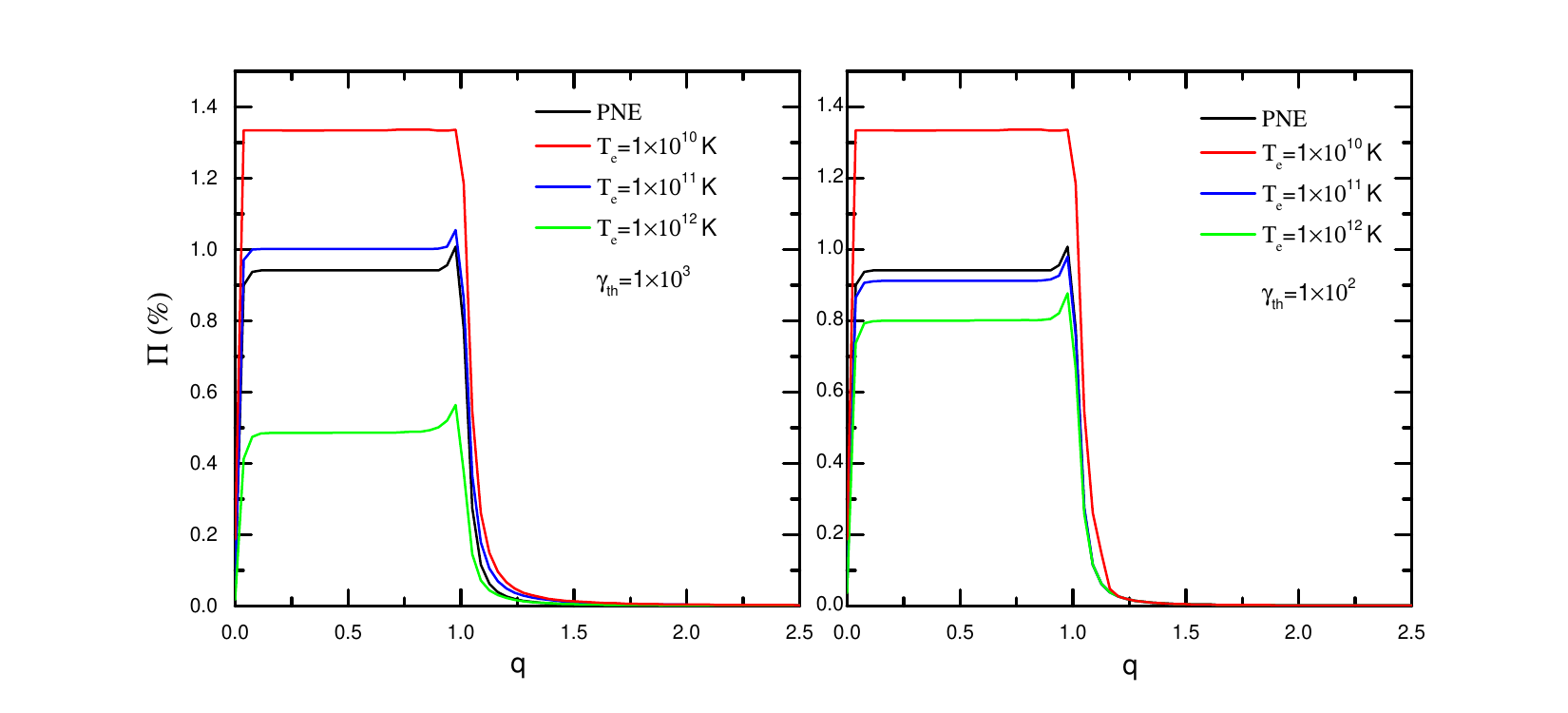}
\caption{Time-averaged PDs with different normalized viewing angles ($q$) for various $T_e$ in the optical band ($6\times 10^{14}$Hz). The "PNE" model represents the synchrotron radiation produced by the pure nonthermal electrons. We take $\gamma_{th}=1\times 10^{3}$ in the left panel and $\gamma_{th}=1\times 10^{2}$ in the right panel. \label{fig:int-cgth}}
\end{figure}

In this section, we calculate the time-averaged circular PDs with different normalized viewing angles ($q$) for various electron temperatures $T_e$ in the optical band. These results are shown in Figure \ref{fig:int-cgth}. We take $\gamma_{th}=1\times 10^{3}$ and $\gamma_{th}=1\times 10^{2}$ in the left panel and in the right panel, respectively. The profiles of the circular PDs with $q$ for different electron temperatures and the profiles for the PNE model are similar. The circular PDs are approximately constant when observing on-beaming ($q\lesssim 1$), while observing off-beaming ($q>1$), the circular PDs drop sharply, by two to three orders of magnitude compared to the on-beaming case. This makes the circular polarization difficult to be detected when observing off-beaming.
Moreover, the lower the temperatures the higher the time-averaged circular PDs in the optical band. Compared to the PNE model, the PDs could be higher than those of the PNE model for the electron temperatures as low as $\sim 10^{10}$ K. The time-averaged circular PD in the hybrid electron model is higher than $\sim 1\%$ when the electron temperature is as low as $\sim 10^{10}$ K, while in the pure nonthermal electron model is lower than $\sim 1\%$.

\begin{figure}[ht!]
\plotone{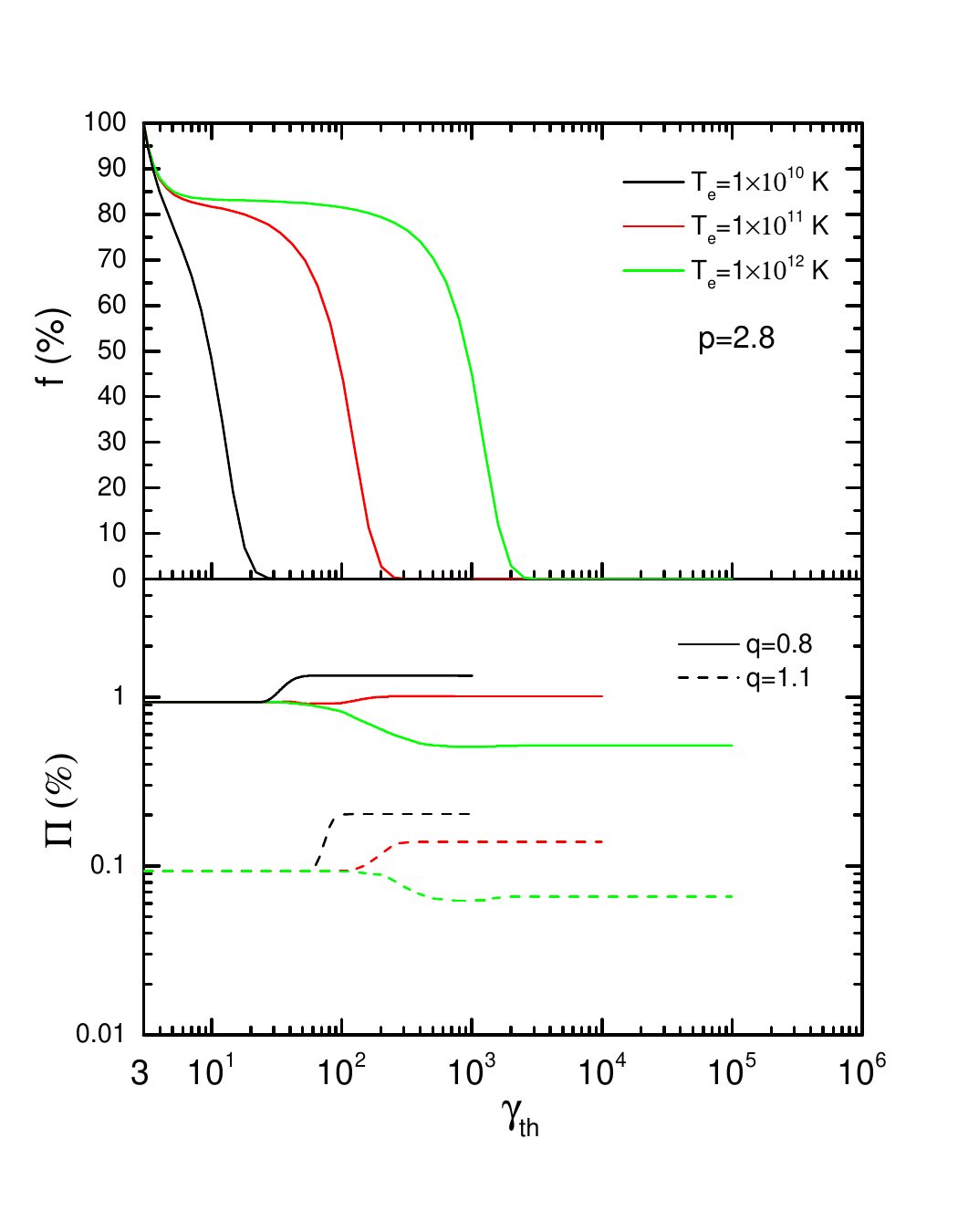}
\caption{Time-averaged circular PDs and the nonthermal energy fraction profile with different $\gamma_{th}$ in various $T_e$ in the optical band. Upper panel: The nonthermal energy fraction with different $\gamma_{th}$. Bottom panel: The time-averaged circular PDs with different $\gamma_{th}$. \label{fig:int-cTe}}
\end{figure}

\subsubsection{Time-averaged circular PD profiles with conjunctive LFs} \label{subsec:general}

In order to further study the relations of circular PD and the conjunctive LFs $\gamma_{th}$ and electron temperatures $T_e$, we calculate the time-averaged circular PDs with different $\gamma_{th}$ in various $T_e$ in the optical band. To understand the effect of electron composition on the circular polarization, we also give the distribution of nonthermal energy fraction $f$ with the conjunctive LFs. The results are shown in Figure \ref{fig:int-cTe}. According to the results, the nonthermal energy fraction has an important effect on the circular PDs. For the electron temperatures $\gtrsim 10^{12}$K, when the nonthermal energy fraction drops to a very low level, the circular PDs will quickly drop to a lower PD platform. On the contrary, the circular PDs will quickly rise to a higher PD platform when the nonthermal energy fraction drops to a low level for the electron temperatures $\sim 10^{10}$ K. When $\gamma_{th} \lesssim 30$, the circular PDs at different electron temperatures are exactly the same. This is because the radiation in the optical band ($6\times 10^{14}$Hz) is mainly generated by electrons with $\gamma_{th}\sim 30$. Thus, when $\gamma_{th} \lesssim 30$, the radiation in the optical band is dominated by the nonthermal electrons, and the circular PDs are not affected by thermal electron temperatures. While $\gamma_{th} \gtrsim 30$, the radiation in the optical band is dominated by the thermal electrons, and the circular PDs would be affected by the electron temperatures. Moreover, we can also draw the conclusion that the lower the electron temperature, the higher the circular PD for a certain conjunctive LF.

\section{Synchrotron polarization radiative transfer in prompt optical emission}\label{sec:transfer}
\begin{figure}[ht!]
\plotone{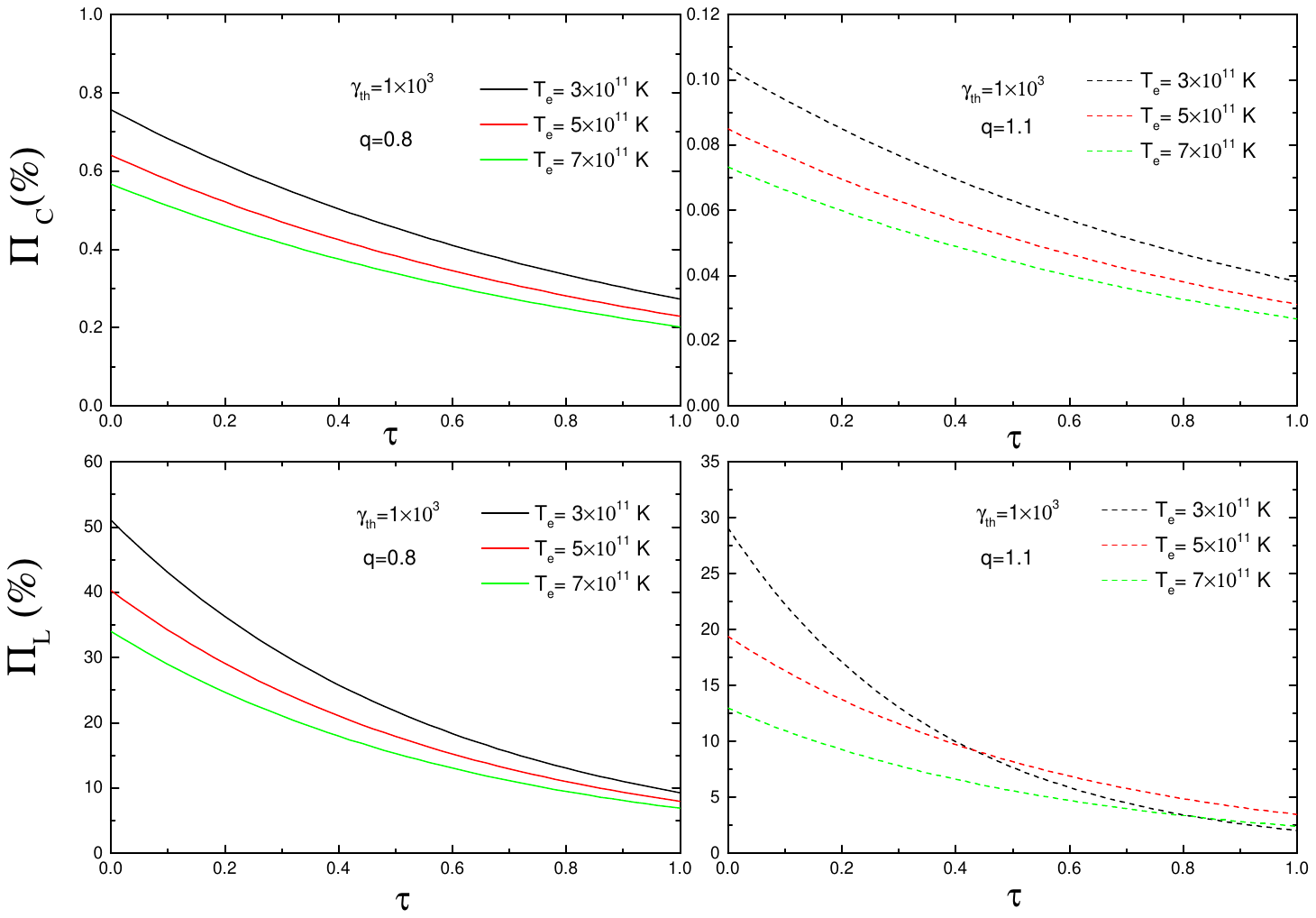}
\caption{The circular and linear PDs profile with optical depth in various electron temperatures. Upper panels: The circular PDs profile with optical depth; Bottom panels: The linear PDs profile with optical depth. Note that $\gamma_{th}=1\times 10^3$ in this group of calculations. \label{fig:tau-pol}}
\end{figure}

The calculations in section \ref{subsec:time-resolve} and \ref{subsec:averaged} have not considered the polarization radiative transfer. In general, polarization radiative transfer should be taken into account in the optical band. The synchrotron polarization radiative transfer have been investigated \citep{Sazonov+1969, Jones+ODell+1977, Jones+Hardee+1979}. 
\cite{Mao+etal+2018} considered the hybrid energy distribution of relativistic thermal and nonthermal electrons and calculated the linear and circular polarization radiative transfer. However, the calculations in Mao18 have not considered the geometric effect of GRB jet. In this paper, we follow the calculations of Mao18 and take into account the jet geometric effect to calculate the synchrotron polarization radiative transfer in the prompt optical emission. The absorption coefficient of the Stokes parameter $I$ is written as 
\begin{equation}\label{kI}
\begin{aligned}
k_I &= \frac{-\sqrt{3} e^2}{4\pi mc\nu^2} \int_{0}^{(1+q)^{2}y_{j}} g(y) dy \int_{-\Delta \phi(y)}^{\Delta \phi(y)} d\phi  \int_{\gamma_{\rm{min}}}^{\gamma_{\rm{max}}} \gamma^2 \nu_L \frac{\partial}{\partial \gamma} [\frac{N_{e}(\gamma)}{\gamma^2}] \\
&\times \sin \alpha' [x\int_{x}^{\infty}K_{5/3}(\xi)d\xi] d\gamma 
\end{aligned}
\end{equation}
The absorption coefficient of the Stokes parameter $Q$ is
\begin{equation}\label{pi}
\begin{aligned}
k_Q &= \frac{-\sqrt{3} e^2}{4\pi mc\nu^2} \int_{0}^{(1+q)^{2}y_{j}} g(y) dy \int_{-\Delta \phi(y)}^{\Delta \phi(y)} d\phi  \int_{\gamma_{\rm{min}}}^{\gamma_{\rm{max}}} \gamma^2 \nu_L \frac{\partial}{\partial \gamma} [\frac{N_{e}(\gamma)}{\gamma^2}] \\
&\times \sin \alpha' [xK_{2/3}(x)] d\gamma 
\end{aligned}
\end{equation}
The absorption coefficient of the Stokes parameter $V$ is
\begin{equation}\label{pi}
\begin{aligned}
k_V &= \frac{-e^2}{\sqrt{3} \pi mc\nu^2} \int_{0}^{(1+q)^{2}y_{j}} g(y) dy \int_{-\Delta \phi(y)}^{\Delta \phi(y)} d\phi  \int_{\gamma_{\rm{min}}}^{\gamma_{\rm{max}}} \gamma \nu_L \frac{\partial}{\partial \gamma} [\frac{N_{e}(\gamma)}{\gamma^2}] \\
&\times \cos \alpha' [xK_{1/3} (x) + \int_{x}^{\infty}K_{1/3}(\xi)d\xi] d\gamma 
\end{aligned}
\end{equation}
The coefficients of rotativity and convertibility are
\begin{equation}\label{pi}
\begin{aligned}
k_{V}^{*} &= - \frac{2 e^2}{mc\nu^2} \int_{0}^{(1+q)^{2}y_{j}} g(y) dy \int_{-\Delta \phi(y)}^{\Delta \phi(y)} d\phi  \int_{\gamma_{\rm{min}}}^{\gamma_{\rm{max}}} \nu_L (\gamma ln \gamma) \frac{\partial}{\partial \gamma} [\frac{N_{e}(\gamma)}{\gamma^2}] \\
&\times \cos \alpha'd\gamma 
\end{aligned}
\end{equation}
\begin{equation}\label{pi}
\begin{aligned}
k_{Q}^{*} &= - \frac{3^{1/3} e^2}{2^{4/3} mc\nu^2} \int_{0}^{(1+q)^{2}y_{j}} g(y) dy \int_{-\Delta \phi(y)}^{\Delta \phi(y)} d\phi  \int_{\gamma_{\rm{min}}}^{\gamma_{\rm{max}}} x \nu_L \gamma^2 \frac{\partial}{\partial \gamma} [\frac{N_{e}(\gamma)}{\gamma^2}] \\
&\times \sin \alpha' \{ \int_{0}^{\infty} \xi \cos [\xi (3\nu/2\nu_c)^{2/3} + \xi^3 /3]d\xi \} d\gamma 
\end{aligned}
\end{equation}
 The radiation transfer matrix can be presented as \citep{Mao+etal+2018}
\begin{equation}\label{matrix}
\begin{bmatrix}
  d/dl + k_I & k_Q & 0 & k_V\\
  k_Q & d/dl+k_I & k_{V}^{*} & 0 \\
  0 & -k_{V}^{*} & d/dl + k_I & k_{Q}^{*} \\
  k_V & 0 & -k_{Q}^{*} & d/dl + k_I
\end{bmatrix} \times
\begin{bmatrix}
 I\\
 Q \\
 0 \\
 V
\end{bmatrix} =
\begin{bmatrix}
 0\\
 0 \\
 0 \\
 0
\end{bmatrix} 
\end{equation}
The Stokes parameter $U$ is assumed to be zero. We solve the 
Equation \ref{matrix} and obtain the circular and linear PDs. The circular PD can be written as

\begin{equation}
\Pi_{C}=\frac{V}{I}=\Pi_{C,0} \exp{(K_C \tau)}
\end{equation}
The linear PD is
\begin{equation}
\Pi_{L}=\frac{Q}{I}=\Pi_{L,0} \exp{(K_L \tau)}
\end{equation}

where $\tau$ is optical depth and $d\tau=k_I dl$. $\Pi_{C,0}$ and $\Pi_{L,0}$ are the circular and linear PDs without polarization radiative transfer, respectively. $\Pi_{C,0}$ is calculated by Equation \ref{pi-0}. $\Pi_{L,0}$ can be calculated as \citep{cheng+etal+2024b}
\begin{equation} \label{pi-nu}
\begin{aligned}
\Pi_{L,0} (\nu) &=\frac{Q(\nu)}{I(\nu)}= \int_{0}^{(1+q)^{2}y_{j}} g(y) dy \int_{-\Delta \phi(y)}^{\Delta \phi(y)} d\phi  \int_{\gamma_{\rm{min}}}^{\gamma_{\rm{max}}} xK_{2/3}(x) N_{e}(\gamma) \\
&\times B'(t') \sin \alpha' \cos (2\chi ) d\gamma  [ \int_{0}^{(1+q)^{2}y_{j}} g(y) dy\\
&\times \int_{-\Delta \phi(y)}^{\Delta \phi(y)} d\phi  \int_{\gamma_{\rm{min}}}^{\gamma_{\rm{max}}} x\int_{x}^{\infty}K_{5/3}(\xi)d\xi N_{e}(\gamma) B'(t')  \sin \alpha'  d\gamma]^{-1}.
\end{aligned}
\end{equation}
The parameters $K_{C}$ and $K_{L}$ determine the distribution of circular and linear PDs with optical depth, respectively. They are written as
\begin{equation}
\begin{aligned}
K_{C} &=(-1-\frac{k_V}{k_I} \sqrt{1+\frac{k_Q k_{Q}^{*}}{k_V k_{V}^{*}}}) \\
& - \frac{1}{2} [k_0 -1 - \sqrt{(1-k_0)^2-4(1-\frac{k_0 k_V}{k_{I}^{2}})}]
\end{aligned}
\end{equation}
and
\begin{equation}
\begin{aligned}
K_{L} &=(-1-\frac{k_Q}{k_I} \sqrt{1+\frac{k_V k_{V}^{*}}{k_Q k_{Q}^{*}}}) \\
& - \frac{1}{2} [k_0 -1 - \sqrt{(1-k_0)^2-4(1-\frac{k_0 k_V}{k_{I}^{2}})}]
\end{aligned}
\end{equation}

where $k_0=k_V + k_Q k_{Q}^{*}/k_{V}^{*}$. We take $\gamma_{th}=1\times 10^3$ in this group of calculations. The circular and linear PDs profile with optical depth in various electron temperatures are shown in Figure \ref{fig:tau-pol}. As is shown, both of the circular and linear PDs decrease with the increase of optical depth, but the linear PDs decline faster than the circular PDs. Thus the influence of radiation transfer on linear polarization is greater than that on circular polarization. The attenuation rates of the PDs with optical depth differs across temperatures, and at a given temperature, the linear PDs attenuate more rapidly than the circular PDs. Moreover, the lower the electron temperatures the faster the circular and linear PDs decrease with the optical depth. Hence, the lower the electron temperatures, the greater the impact of radiation transfer in circualr and linear PDs.

\section{Summary and discussion} \label{sec: conclusion}
We calculate the synchrotron circular PD of the hybrid distribution of relativistic thermal and nonthermal electrons with different viewing angles ($q$), electron temperatures ($T_e$), and conjunctive LFs ($\gamma_{th}$) in GRB prompt optical emission, including instantaneous and time-averaged circular PDs. Furthermore, we calculate synchrotron polarization radiative transfer in GRB prompt optical emission, including circular and linear polarization radiative transfer. Our main results are summarized as follows.

\begin{enumerate}

\item The lower the thermal electron temperature, the higher the circular PD in the prompt optical emission. The circular PD calculated by the hybrid electron distribution model in the case of the electron temperature as low as $\sim 10^{10}$ K is higher than that calculated by the pure non-thermal electron model. It makes the circular polarization of the hybrid electron model to be relatively easier to detect.

\item  Comparing to the case of solely nonthermal electrons, the synchrotron circular PDs in the hybrid electron model can vary widely, depending on different parameters ($q$, $T_e$, and $\gamma_{th}$).

\item  Both of the circular and linear PDs decrease with the increase of optical depth, but the linear PDs decline faster than the circular PDs. Moreover, the lower the electron temperatures the faster the circular and linear PDs decrease with the optical depth.

\item In the uniform top-hat jet model, the circular PD is very low for the off-beaming case. This makes the circular polarization difficult to detect when observing off-beaming.
\end{enumerate}

Under the hybrid electron energy distribution model, the time-averaged circular PDs can be higher than $\sim 1\%$ when the electron temperature is as low as $\sim 10^{10}$ K, while in the solely nonthermal electron model usually lower than $\sim 1\%$. Therefore, the observation of time-averaged circular PD higher than $\sim 1\%$ in the prompt optical emission may indicate the presence of relativistic thermal electrons. Furthermore, combined with the results on synchrotron linear polarization presented in \cite{cheng+etal+2024b} that the observed GRBs with linear PD higher than $\sim 60\%$ in the gamma-ray or X-ray bands may also indicate
the existence of relativistic thermal electrons. If a GRB exhibits both circular and linear polarizations that satisfy the above two criteria simultaneously, we can consider this GRB likely to be partially radiated from relativistic thermal electrons. These results can be jointly verified through multi-wavelength polarization observations, including polarization measurements in the optical, X-ray, and gamma-ray bands. The optical polarization can be measured by polarimeters such as MOPTOP \citep{Shrestha+etal+2020}, while X-ray and gamma-ray polarization can be measured by eXTP \citep{eXTP2019} and POLAR-2 \citep{Kole+etal+2024, Feng+etal+2024}, respectively. This will provide new insights into particle acceleration mechanisms in GRB jets.

\begin{acknowledgments}
This work is supported by the Scientific Research Project of Guangxi Minzu University (2021KJQD03), the Guangxi Natural Science Foundation (2024GXNSFBA010350), the Guangxi Natural Science Foundation (GUIKEAD22035945), the National Key R\&D Program of China (2023YFE0101200), and the National Natural Science Foundation of China (Nos. 12393813, 12393811, 12027803). J.M is supported by the Yunnan Revitalization Talent Support Program (YunLing Scholar Project).
\end{acknowledgments}

%





\bibliography{sample631}{}
\bibliographystyle{aasjournal}



\end{document}